# Origin of giant magnetoresistance in layered nodal-line semimetal TaNiTe$_5$ nanoflakes

Ding-Bang Zhou,[1] Kuang-Hong Gao,[1,*] Meng-Fan Zhao,[2] Zhi-Yan Jia,[2] Xiao-Xia Hu,[3] Qian-Jin Guo,[3] Hai-Yan Du,[3] Xiao-Ping Chen,[3] and Zhi-Qing Li[1,†]

[1]*Tianjin Key Laboratory of Low Dimensional Materials Physics and Preparing Technology, Department of Physics, Tianjin University, Tianjin 300354, China*

[2]*Institute of Quantum Materials and Devices, Tiangong University, Tianjin 300387, China*

[3]*Analysis and Testing Center of Tianjin University, Tianjin University, Tianjin 300072, China*

**Abstract**

Layered transition metal chalcogenides have stimulated a wide research interest due to its many exotic physical properties. In this paper, we studied the magnetotransport properties of the exfoliated nodal-line semimetal TaNiTe$_5$. A giant positive magnetoresistance (MR) is observed when the current is parallel to the crystallographic *c* axis, while it is strongly diminished when the current flows along the *a* axis. The observed giant MR is gradually suppressed either on reducing the thickness of nanoflake or on increasing temperature. By performing MR measurement in tilted magnetic fields, the interlayer coupling is found to be weakened both by reducing the thickness and by increasing temperature. We propose a mechanism of electron-electron interaction-assisted interlayer transport as a origin of the giant MR. The mechanism is likely to provide an explanation for the giant MR in other layered materials.

---

[*] Corresponding author, e-mail: khgao@tju.edu.cn
[†] Corresponding author, e-mail: zhiqingli@tju.edu.cn

# I. INTRODUCTION

Layered transition metal chalcogenides have revealed a range of unique electronic phenomena, including two-dimensional superconductivity [1], charge density waves [2], anisotropic chiral anomaly [3], and coexistence of ferroelectric polarization and Dirac surface state [4]. This makes them have a great potential both for fabricating electronic devices and for fundamental research to explore exotic quantum phases. In 2014, $WTe_2$, as a typical transition metal chalcogenide, was reported to exhibit an extremely large positive magnetoresistance (MR) effect, which can be as high as 13 million percent and has potential applications in magnetic sensors and memories [5]. This finding has stimulated extensive research interest in transition metal chalcogenides. Although the extremely large positive MR effect has been observed in several other transition metal chalcogenides such as $MoTe_2$ and $ZrTe_2$ [6,7], its origin is still under intensive debate. Electron-hole compensation is usually used to explain large MR effect. However, Wang *et al*. claimed that the electron-hole compensation is not the mechanism for large MR in $WTe_2$ [8]. Similarly, in thin $WTe_2$ and $MoTe_2$, the magnitude of MR was found to be determined by high carrier mobility rather than by the electron-hole compensation [9,10]. Alternatively, the topological property of carriers was reported to play a dominant role in the occurrence of large MR in $WTe_2$ and $SnTaS_2$ [11,12]. Thus, it is necessary to further investigate the MR effect in layered transition metal chalcogenides.

Recently, it has been found that $TaNiTe_5$ compound, a member of the transition metal chalcogenide family, is a Dirac nodal-line semimetal with fourfold degeneracy [13]. A giant positive MR was found in the compound, but its origin has not been fully explored [14]. On the other hand, regarding to the dimensionality of $TaNiTe_5$, it was theoretically predicted to be a two-dimensional topological semimetal [15,16]. However, Xu *et al*. found that $TaNiTe_5$ hosts quasi-one-dimensional topological electronic properties via magnetization and electrical transport measurements [17]. On the contrary, Chen *et al*. performed magnetotransport and de Hass−van Alphen effect studies and suggested that $TaNiTe_5$ is a three-dimensional topological semimetal [18]. Therefore, the magnetotransport property of $TaNiTe_5$ needs to be further studied.

In this paper, we systematically performed magnetotransport measurements of TaNiTe$_5$ nanoflakes with different thicknesses. A giant positive MR is observed when an applied current is along crystallographic $c$ axis, while it is strongly diminished when the current is along the $a$ axis. The observed giant MR is gradually suppressed with reducing thickness and increasing temperature. Regarding its origin, we found that the observed MR cannot be explained by weak antilocalization or by electron-hole compensation. We attribute it to the interlayer transport assisted by electron-electron interaction.

## II. EXPERIMENTAL METHOD

We fabricated high quality TaNiTe$_5$ single crystal by chemical vapor transport method. Powders of Ta (99.998%), Ni (99.999%), and Te (99.999%) with a 1:1:5 molar ratio were mixed and sealed in a quartz tube under high vacuum. The quartz tube was put into a furnace and heated from room temperature up to 973 K within 24 h. Then the temperature was maintained for 4 days, and slowly cooled to 773 K within 24 h. Last, the quartz tube was taken out from the furnace for natural cooling to room temperature. The grown TaNiTe$_5$ single crystals were characterized by X-ray diffraction (XRD), transmission electron microscopy (TEM), and energy dispersive X-ray spectroscopy (EDS) measurements. TaNiTe$_5$ nanoflakes with different thicknesses were obtained by the mechanical exfoliation of the grown single crystals. The exfoliated nanoflakes were transferred onto silicon substrates with a 300 nm SiO$_2$ cover layer. Their thicknesses were determined by atomic force microscopy (see Table 1). For the nanoflakes, their exposed surfaces were found to be the crystallographic $ac$ plane (i.e., normal to the $b$ axis). The measured patterns were defined by a standard photolithography. For ribbon-shaped nanoflakes, four Ohmic contacts were deposited by thermal evaporation of Ti/Au [see the insets of Figs. 2(a) and (b)], while six Ohmic contacts were deposited for the wider nanoflakes [see the insets of Figs. 2(c) and (d)]. Five nanoflakes studied in this paper can be classified into two groups. One group contains two samples with different thicknesses (referred to as samples S1 and S2, respectively), in which the applied current is along the $a$ axis. The other group contains three samples (referred to as samples S3, S4, and S5, respectively), in which the applied current is along the $c$ axis. The magnetotransport properties were measured in a physical property measurement system (PPMS, Quantum Design).

### III. RESULTS AND DISCUSSION

Figure 1(a) shows the XRD pattern of the grown TaNiTe$_5$ single crystal. All observed peaks can be indexed as (0$L$0) reflections of TaNiTe$_5$ and no secondary phase is detected. From the positions of these peaks, the lattice parameter $b$ was calculated to be equal to 13.09 Å, consistent with the reported value of 13.17 Å [19]. Figure 1(b) shows a high-resolution TEM image of the grown TaNiTe$_5$. One can see a well-defined atomic ordered arrangement that is indicative of high crystalline quality. From the EDS measurement, the atomic ratio of Ta:Ni:Te is determined to be 0.95:1.06:5.13 that is near to 1:1:5 for the stoichiometric TaNiTe$_5$. The EDS images for three elemental Ta, Ni, and Te, as shown in Fig. 1(c), demonstrate that all of them are uniform distributions without secondary phase, consistent with the XRD data.

Figure 2(a) shows longitudinal resistivity $\rho_{xx}$ as a function of temperature $T$ for sample S1 under different magnetic field $B$s. Here $B$ is perpendicular to the plane of sample. For $B = 0$ T, one can see that $\rho_{xx}$ exhibits a decrease with $T$ decreasing from 300 down to ~10 K. This is suggestive of a typical metallic behavior. As $T$ is further reduced to the lowest measured temperature of 2 K, $\rho_{xx}$ becomes saturated and forms a plateau at low $T$. We can define residual resistivity ratio (RRR) as $\rho_{xx}(300\text{ K}) / \rho_{xx}(2\text{ K})$. It was calculated to be 29.1, comparable with the reported result in bulk TaNiTe$_5$ [18]. $\rho_{xx}$ in the low $T$ range of 2–30 K can be fitted by using a relation $\rho_{xx} = A + CT^2$ ($A$ and $C$ are two constants), indicating a Fermi liquid behavior in which electron-electron interaction (EEI) dominates electrical transport. As $B$ increases, the $\rho_{xx}$ plateau demonstrates a vertical upward shift, but no visible $B$ dependence for $\rho_{xx}$ is observed at high $T$. The upward shift of $\rho_{xx}$ plateau with $B$ indicates a presence of positive MR at low $T$. A similar phenomenon is also observed in sample S2 (not given here). Figure 2(b) shows $\rho_{xx}$ as a function of $T$ for sample S3 under different $B$s. For $B = 0$ T, similarly to sample S1, $\rho_{xx}$ shows a typical metallic character at high $T$, accompanied by the appearance of plateau at low $T$. The EEI also dominates electrical transport at low $T$ (dashed line is a fit according to $\rho_{xx} = A + CT^2$). When $B > 0\,\mathrm{T}$, compared with sample S1, there are two characteristics in these curves to note. First, the vertical shift of $\rho_{xx}$ plateau with $B$ is more remarkable, in agreement with the reported results [17]. This is indicative of a larger positive MR when the current flows along the $c$ axis. Second, for $B > 4\,\mathrm{T}$, $\rho_{xx}$ shows an increase with decreasing $T$ in the intermediate $T$ range of 10– ~30 K. That is, we observe so-called "turn-on" behavior that has been widely found in other layered transition metal chalcogenides. When the thickness of nanoflakes is reduced, as shown in Fig. 2(c) for sample S4, the vertical shift of $\rho_{xx}$ plateau with $B$ seems to remain unchanged. But the "turn-on" behavior disappears when the thickness of nanoflake is further reduced to 33.7 nm for sample S5 [see Fig. 2(d)].

Figure 3(a) shows the MR curves for all our samples at 2 K. Here, the MR is defined as $\mathrm{MR} = [\rho_{xx}(B) - \rho_{xx}(0)] / \rho_{xx}(0)$. One can see that two MR curves for the first group of samples S1 and S2 are almost overlapped, with a maximum value of 43% under $B = 8\,\mathrm{T}$. This indicates that the MR is thickness independent when the current is along the $a$ axis. Interestingly, the magnitude of MR is far smaller than that in the second group of samples S3–S5 with the current along the $c$ axis. As seen in Table 1, MR varies between 132.5–486.8% at 2 K and 8 T for the second group. This implies that there is an obvious current direction dependence for MR in TaNiTe$_5$. Strikingly, the MR in the second sample group exhibits a strong thickness dependence: it is gradually suppressed with reducing the thickness of nanoflakes for samples S3-S5 in sequence. This is completely different from observations in the first group of samples S1 and S2, implying that there must be different MR origins between two sample groups. The giant MR in samples S3–S5 is of interest because of potential applications. Figure 3(b) shows a $T$ dependence of MR in sample S3. One can see that the observed MR is gradually diminished as $T$ increases, and it is invisible above 50 K. A plateau of MR at low $T$ is observed for $B = 8\,\mathrm{T}$, as seen in the inset of Fig. 3(b).

What causes the giant MR in samples S3–S5? Weak antilocalization (WAL) effect arising from an additional $\pi$ Berry phase of nodal line is a possible source [11,12]. In order to check this possibility, we fitted magnetoconductivity [defined as $\Delta\sigma_{xx}(B) \approx 1/R_s(B) - 1/R_s(0)$, where $R_s$ is sheet resistance] using two-dimensional Hikami-Larkin-Nagaoka (HLN) theory in a strong limit of spin-orbit coupling [20]:

$$\Delta\sigma_{xx}(B) = \alpha\frac{e^2}{\pi h}\left[\ln\left(\frac{B_\varphi}{B}\right) - \Psi\left(\frac{1}{2} + \frac{B_\varphi}{B}\right)\right], \tag{1}$$

where $\alpha$ is a factor, $e$ is electronic charge, $h$ is Plank's constant, $B_\varphi = h/(8\pi e L_\varphi^2)$ ($L_\varphi$ is phase coherence length), $\Psi(x)$ is digamma function. As shown in Fig. 3(c) for sample S3 as a representative example, the fitting curve follows the experimental data very closely in the low $B$ range of $|B| < 3$ T for any fixed $T$. The extracted $L_\varphi$ vs $T$ curve is given in Fig. 3(d). One can see that the data can be described by the relation $L_\varphi \sim T^{-0.5}$. This is a further evidence that there exists EEI that dominates dephasing process for carriers [21,22]. The presence of deviation below ~15 K can be attributed to the saturation term [23]. The extracted $\alpha$ from the fits is also given in Fig. 3(d). We note that it is very large and exhibits a slight decrease with increasing $T$. Very large $\alpha$ value may result from the bulk conduction channels [12,24,25]. More importantly, as seen in Fig. 3(c), there appears a large deviation between fitting curve and experimental data at high $B$. This manifests that the observed giant MR in samples S3–S5 cannot be attributed to the WAL effect. Furthermore, considering the WAL effect arising from the topological protected carriers, it is reasonable to deduce that the giant MR in samples S3–S5 is not related to the presence of Dirac point. This is consistent with observation in layered semimetal HfTe$_2$ [26]. On the other hand, a disorder-induced nonsaturating linear MR is widely reported in other layered materials such as PtBi$_2$ and Bi$_2$Se$_3$ [27,28]. But the giant MR in our samples cannot be attributed to the disorder due to two reasons. First, the observed giant MR displays a quasi-quadratic field dependence [Fig. 3(a)], which is not a linear field dependence. Second, the *RRR* magnitude of samples S3–S5 is thickness independent (see Table 1), signifying that the disorder is not significantly changed with reducing the thickness. If the giant MR arose from the disorder, it should be thickness independent, either. Contrarily, the giant MR in the second sample group exhibits the strong thickness dependence [Fig. 3(a)]. This proves that the giant MR in our samples cannot be explained by the disorder.

Another possible origin is the electron-hole compensation that is a prevailing explanation for large MR effect in transition metal chalcogenides. Theoretically, a perfect electron-hole compensation will yield a nonsaturating MR with a quadratic field dependence [29]. In order to clarify the contribution of electron-hole compensation to MR, we adopt a two-band model [30] where the $B$ dependences of $\rho_{xx}$ and Hall resistivity $\rho_{xy}$ are given as

$$\rho_{xx}(B) = \frac{(n\mu_n + p\mu_p) + (n\mu_p + p\mu_n)\mu_n\mu_p B^2}{e[(n\mu_n + p\mu_p)^2 + (p-n)^2 \mu_n^2 \mu_p^2 B^2]},\tag{2}$$

$$\rho_{xy}(B) = \frac{(p\mu_p^2 - n\mu_n^2)B + (p-n)\mu_p^2 \mu_n^2 B^3}{e[(n\mu_n + p\mu_p)^2 + (p-n)^2 \mu_n^2 \mu_p^2 B^2]},\tag{3}$$

where $n$ ($p$) and $\mu_n$ ($\mu_p$) are electron (hole) concentration and mobility, respectively. Since Hall measurement can be realized in samples S4 and S5, we tried to perform a simultaneous fit of both $\rho_{xx}$ and $\rho_{xy}$ data in these two samples. Unfortunately, we found it impossible to obtain a good fit, presumably due to the presence of aforementioned WAL at low $B$ that can complicate the fit. Hence, only $\rho_{xy}$ vs $B$ curve is analyzed to obtain carrier concentration and mobility. Figure 4 (a) shows $\rho_{xy}$ as a function of $B$ at various $T$s for sample S4. One can see that the slopes of these curves are positive for any fixed $T$. This implies that hole plays a dominant role, in agreement with the reported result [13]. Importantly, there is a slightly nonlinear $B$ dependence for $\rho_{xy}$ especially below 100 K. This signifies a coexistence of multiple carriers, in agreement with the theoretical prediction [see the inset of Fig. 4(c)][17]. We fit these $\rho_{xy}$ vs $B$ curves by using Eq. (3). As can be seen in Fig. 4 (a), Equation (3) describes the experimental data very well at various $T$s . The fit-obtained $\mu_n$ and $\mu_p$ are given in Fig. 4(b). Both of them decrease slowly on increasing $T$, which can be attributed to electron-electron and/or electron-phonon scattering. The fit-obtained carrier concentrations are given in Fig. 4(c). Both $n$ and $p$ are $T$ independent below 30 K. But the latter displays a remarkable increase at high $T$, presumably resulting from thermal activation. Similar phenomenon is also reported in TaNiTe$_5$ single crystal [14]. It is worth noting that $p$ value is far larger than $n$ value, leading to a large ratio $p/n = 13.9$ below 30 K (a larger value is expected at high $T$). The large ratio implies a breakdown of electron-hole compensation in our samples, proving that it is not the origin of the observed giant MR. To further confirm this point, the MR was further calculated by substituting the fit-obtained values into Eq. (2). As shown in Fig. 4(d) for sample S4 at 2 K as a representative example, we find that the magnitude of calculated MR is far smaller than that of the measured MR (quantitatively, for $B = 8$ T, the calculated MR is only 42.4% while the measured value is as high as 408.3%). Furthermore, the calculated MR tends to saturate above 6 T, which is completely different from the shape of the measured MR. This, combined with the smaller magnitude of the calculated MR, supports that the electron-hole compensation cannot be used to explain the giant MR in samples S3-S5.

A $B$-induced energy gap [31,32] has been proposed to explain the $\rho_{xx}$ "turn-on" behavior [Figs. 2(b) and (c)] that might be related to the giant MR. However, the normalized MR versus $T$ curves at various $B$s, as seen in the inset of Fig. 5(a) for sample S3, converges on a single curve, indicating a same $T$ dependence. This manifests that the $B$-induced energy gap does not exist in our samples [33]. On the other hand, Kohler's rule was believed to provide a good explanation for the $\rho_{xx}$ "turn-on" behavior in WTe$_2$ and MoTe$_2$ [34,35]. Figures 5(a)–(c) show the MR vs $B/\rho_0$ curves (here, $\rho_0$ is longitudinal resistivity under $B = 0$ T) at various $T$s for samples S3–S5. As can be seen in these figures, all MR data for a given sample is scaled onto a single curve. This indicates that the $T$ dependence of MR follows the Kohler's rule

$$\text{MR} = \beta(B/\rho_0)^m, \tag{4}$$

where both $\beta$ and $m$ are constants. We have used Eq. (4) to fit the MR vs $B/\rho_0$ curves [red dashed lines are fits in Figs. 5(a)–(c)]. The fit-obtained $m$ varies between 1 and 2 (namely, $m < 2$), deviating from the perfect electron-hole compensation with $m = 2$ [36]. This is further evidence that the electron-hole compensation is broken in our samples, in agreement with the analysis of Hall data in Fig. 4. Equation (4) can be rewritten as

$$\rho_{xx}(T, B) = \rho_0 + \beta B^m / \rho_0^{m-1}. \tag{5}$$

Combined the relation $\rho_0 = A + CT^2$ at low $T$ (see Fig. 2), Equation (5) perfectly reproduces $\rho_{xx}$ vs $T$ curves under different $B$s, as shown in Fig. 5(d) for a representative sample S3. This indicates that Kohler's rule indeed provides a good explanation for the $\rho_{xx}$ "turn-on" behavior. The observations in Figs. 5(a)–(c) that MR follows Kohler's rule only suggest that one single relaxation time dominates the scattering process of carriers, in line with the breakdown of electron-hole compensation. But this cannot uncover the origin of giant MR.

In order to find the origin, one should note three characteristics of giant MR in our samples: (1) it is present when the applied current flows along the $c$ axis while it is strongly diminished when the current is along the $a$ axis, (2) it is progressively suppressed both as the thickness of nanoflake is reduced and as $T$ increases, and (3) its appearance is accompanied by the EEI at low $T$. Kastrinakis theoretically proposed that the giant MR arises from the combination of elastic spin disorder scattering and a special value of the Hubbard constant [37,38]. But this mechanism cannot explain the characteristic (1) of giant MR in our samples, indicating that it can be negligible. The characteristic (1) must be closely related to the crystal structure. Theoretically, TaNiTe$_5$ has been reported to have an anisotropic Fermi surface with different valleys in different crystallographic directions [13,17]. It might be a possible source for the observed MR if the characteristic (1) was only considered. We note that there is no significant difference for carrier concentrations in both samples S4 and S5. This is suggestive of comparable Fermi energy in these two samples. But a remarkably suppressed MR is observed in sample S5, excluding the anisotropic Fermi surface with different valleys as the source of giant MR. Furthermore, it cannot offer a reasonable explanation for the $T$ dependence of giant MR, either. We therefore conclude that the observed giant MR cannot be attributed to the anisotropic Fermi surface with different valleys because of the characteristic (2).

Figure 6(a) shows the *ac* plane (normal to the *b* axis) of TaNiTe$_5$ crystal structure. One can see that there are many parallel one-dimensional NiTe$_2$ chains that are stretched along the *a* axis. Two adjacent NiTe$_2$ chains are linked via Ta atoms, which makes these NiTe$_2$ chains form a two-dimensional layer in the *ac* plane. The formed two-dimensional layers, as shown in Fig. 6(b), are stacked along the *b* axis via a weak van der Waals (vdW) force. It is found that transport electrons mainly originate from the NiTe$_2$ chains, while Ta atoms have little contribution to electrical transport [13]. Apparently, a lower electivity can be obtained when the current is along the direction of NiTe$_2$ chains (i.e., along the *a* axis), compared to other current directions. This explains a recent experimental observation that the bulk TaNiTe$_5$ demonstrates a highly anisotropic transport behavior [17]. Then one question arises: what will happen when the applied current is along the other directions such as the *c* axis in our samples S3–S5? Naturally, it is unavoidable that there occurs an inter-chain transport via Ta atoms and/or an interlayer transport across vdW gap.

In Cr$_2$Ge$_2$Te$_6$ thin flakes, the interlayer vdW force was found to cause an anisotropic colossal MR effect [39]. In TaNiTe$_5$ single crystal, there exists quantum oscillations for the in-plane magnetic field (i.e., parallel to the *ac* plane) [18]. Recently, it was reported that there are metallic surface states on the "side-cleaved" surface perpendicular to the vdW layer of TaNiTe$_5$ [40]. All these reported results signify that the interlayer transport across vdW gap has an important influence on the electrical transport behavior of layered compounds.

If the interlayer transport played a dominant part in the presence of giant MR, a reduction in the thickness of nanoflake may weaken the interlayer coupling, which consequently suppresses the giant MR [corresponds to the characteristic (2)]. In order to check this point, the MR effect is further studied in tilted *B*. As shown in the inset of Fig. 7(b), $\theta$ is defined as an angle between *B* and the normal direction of nanoflake (i.e., $\theta = 0^\circ$ in Figs. 2–5). Figure 7(a) shows the MR of a representative sample S3 at 2 K for various $\theta$s. We found that the MR is gradually suppressed with increasing $\theta$ from $0^{o}$ to $90^{o}$. But for $\theta = 90^{o}$, as seen in the inset of Fig. 7(a), the MR does not disappear and its magnitude is 24% under $B = 8 \, \text{T}$. This is suggestive of a three-dimensional anisotropy that is also reported in WTe$_2$ [41], supporting that TaNiTe$_5$ is a three-dimensional semimetal [18]. It can be analyzed by a scaling approach about longitudinal resistance [41,42]

$$R(B, \theta) = R(\varepsilon_\theta B) \tag{6}$$

with a scaling factor

$$\varepsilon_\theta = (\cos^2 \theta + \gamma^{-2} \sin^2 \theta)^{1/2}, \tag{7}$$

where $\gamma^2$ is the ratio of the electron effective mass for $\theta = 0^{o}$ and $90^{o}$. As seen in Fig. 7(b), the MR curves at various $\theta$s are collapsed onto a single curve with a field scaling. From the scaling analysis, we extracted the scaling factor $\varepsilon_\theta$. The $\theta$ dependence of the extracted $\varepsilon_\theta$ for several selected $T$s are shown in Fig. 7(c). As seen in the figure, the experimental data can be well described by Eq. (7) (solid lines are fits). The fit-obtained $\gamma$ as a function of $T$ is given in Fig. 7(d). One can see that $\gamma$ exhibits a rapid decrease from 6.14 to 3.16 on increasing $T$ from 2 to 50 K, and it tends to saturate to 2.23 up to 200 K. This $T$ dependence is presumably due to the thermal expansion of the crystal and/or electron-phonon coupling [43]. These $\gamma$ values are smaller than the reported values of ~12 in graphite [44] and ~9 in superconductor YBa$_2$Cu$_3$O$_7$ [45], but they are comparable to those of 2–5 in WTe$_2$[41], 5.4 in HfTe$_5$ [46], 2.1 in black phosphorus [47]. Most importantly, as shown in the inset of Fig. 7(d), $\gamma$ displays a continuous decrease as the nanoflake is made more thinner for a fixed $T$. This proves that the interlayer coupling is indeed weakened in the thinner samples, as reported in WTe$_2$ nanoflakes [48]. Therefore, the interlayer transport across vdW gap is likely to play a dominant part in the occurrence of the giant MR.

Furthermore, the accompaniment of EEI [i.e., characteristic (3) mentioned above] implies that the EEI may be indispensable for observing the giant MR in our samples. Actually, the EEI has been found to wield influence on the occupation of different bands for carriers [49]. Additionally, the appearance of MR plateau at low $T$ [see the inset of Fig. 3(b)] implies that only a single scattering process of EEI is important for carriers.

Hence, one can speculate that the EEI is closely related to the interlayer coupling in our samples. Considering the dominant role of the interlayer transport, we tentatively propose an alternative mechanism for the giant MR: EEI-assisted interlayer transport. In this mechanism, carriers move across vdW gap between two adjacent $ac$ planes due to EEI. When the current is applied along the direction of $NiTe_2$ chains (i.e., along the $a$ axis), carriers move dominantly along the direction of chains. In this situation, the EEI-assisted interlayer transport can be negligible. As a result, the giant MR is absent. This corresponds to the case of samples S1 and S2, where the observed small MR in Fig. 3(a) may result from the weak antilocalization. In contrast, when the current is applied along the $c$ axis for samples S3–S5, the EEI-assisted interlayer transport dominates the electrical transport behavior of carriers, and then the giant MR appears, as seen in Fig. 3(a). When the nanoflake becomes more thinner, the interlayer coupling is expected to be weakened. The EEI-assisted interlayer transport is thus suppressed to some extent, and consequently the giant MR is diminished in the thinner sample. For a given thickness of nanoflake, the EEI-assisted interlayer transport is also gradually suppressed as $T$ increases because of the interlayer coupling weakness deduced by the decrease in $\gamma$ with $T$ [see Fig. 7(d)]. This explains the observation that the giant MR is diminished as $T$ increases [see Fig. 3(b)]. It can be concluded that our proposed EEI-assisted interlayer transport as a origin can capture all features of the giant MR in our samples.

Importantly, it has been widely reported that the magnitude of large MR is proportional to the thickness of film or nanoflake for other layered materials such as $WTe_2$ [48], $MoTe_2$ [10], $NbTe_2$ [50] and multilayer graphene [51]. Considering the suppression of MR in thinner sample as a typical feature of our proposed mechanism, therefore, one can conjecture that the EEI-assisted interlayer transport-induced MR may be a universal phenomenon in layered compound family. However, it should be mentioned that there is no theoretical model that addresses the proposed scenario and further theoretical study is highly required to quantitatively describe the giant MR. We should emphasize that we do not question the other reasonable interpretations of the giant MR in layered materials, such as disorder effect in layered $HfTe_5$ [46], quantum mechanism in multilayer graphene [52] and perfect electron-hole compensation in $WTe_2$ [5]. Our proposed mechanism can offer a substituting explanation for the giant MR in layered materials.

## IV. CONCLUSION

We study the magnetotransport properties of exfoliated $TaNiTe_5$ nanoflakes with different thicknesses. A small MR with the magnitude less than 50% appears at 2 K and 8 T when the current is along the crystallographic *a* axis. In contrast, a giant positive MR reaching to 486.8% is observed when the current is along the *c* axis. The observed giant MR is gradually suppressed either on reducing the thickness of nanoflake or on increasing temperature. Through the evolution of giant MR with different tilted magnetic fields, the interlayer coupling is found to be weakened both by reducing the thickness and by increasing temperature. The observed MR can be explained not by weak antilocalization or electron-hole compensation but by the EEI-assisted interlayer transport.

## ACKNOWLEDGMENTS

The authors are grateful to Feng Lu, Zi-Wu Wang, and Song-Ci Li for fruitful discussions. This work was supported by the National Natural Science Foundation of China grant No. 12174282.

[1] A. W. Tsen, B. Hunt, Y. D. Kim, Z. J. Yuan, S. Jia, R. J. Cava, J. Hone, P. Kim, C. R. Dean, A. N. Pasupathy, Nature of the quantum metal in a two-dimensional crystalline superconductor, Nat. Phys. **12**, 208 (2015).

[2] X. Xi, L. Zhao, Z. Wang, H. Berger, L. Forro, J. Shan, K. F. Mak, Strongly enhanced charge-density-wave order in monolayer $NbSe_2$, Nat. Nanotechnol. **10**, 765 (2015).

[3] Y.-Y. Lv, X. Li, B.-B. Zhang, W. Y. Deng, S.-H. Yao, Y. B. Chen, J. Zhou, S.-T. Zhang, M.-H. Lu, L. Zhang, M. Tian, L. Sheng, and Y.-F. Chen, Experimental observation of anisotropic Adler-Bell-Jackiw anomaly in type-II Weyl semimetal $WTe_{1.98}$ crystals at the quasiclassical regime, Phys. Rev. Lett. **118**, 096603 (2017).

[4] Y. Li, Z. Ran, C. Huang, G. Wang, P. Shen, H. Huang, C. Xu, Y. Liu, W. Jiao, W. Jiang, J. Hu, G. Zhu, C. Xu, Q. Lu, G. Wang, Q. Jing, S. Wang, Z. Shi, J. Jia, X. Xu , W. Zhang , W. Luo, and D. Qian, Coexistence of ferroelectriclike polarization and Dirac-like surface state in $TaNiTe_5$, Phys. Rev. Lett. **128**, 106802 (2022).

[5] M. N. Ali, J. Xiong, S. Flynn, J. Tao, Q. D. Gibson, L. M. Schoop, T. Liang, N. Haldolaarachchige, M. Hirschberger, N. P. Ong, and R. J. Cava, Large, non-saturating magnetoresistance in $WTe_2$, Nature **514**, 205 (2014).

[6] S. Lee, J. Jang, S.-I. Kim, S.-G. Jung, J. Kim, S. Cho, S. W. Kim, J. Y. Rhee, K.-S. Park, and T. Park, Origin of extremely large magnetoresistance in the candidate type-II Weyl semimetal $MoTe_{2-x}$, Sci. Rep. **8**, 13937 (2018).

[7] H. Wang, C. H. Chan, C. H. Suen, S. P. Lau, and J.-Y. Dai, Magnetotransport properties of layered topological material $ZrTe_2$ thin film, ACS Nano **13**, 6008 (2019).

[8] Y. Wang, K. Wang, J. Reutt-Robey, J. Paglione, and M. S. Fuhrer, Breakdown of compensation and persistence of nonsaturating magnetoresistance in gated $WTe_2$ thin flakes, Phys. Rev. B **93**, 121108(R) (2016).

[9] Y. Yi, C. Wu, H. Wang, H. Liu, H. Li, H. Zhang, H. He, J. Wang, Thickness dependent magneto transport properties of $WTe_2$ thin films, Solid State Commun. **260**, 45 (2017).

[10] S. Zhong, A. Tiwari, G. Nichols, F. Chen, X. Luo, Y. Sun, and A. W. Tsen, Origin of magnetoresistance suppression in thin $\gamma$-MoTe$_2$, Phys. Rev. B **97**, 241409(R) (2018).

[11] J. Jiang, F. Tang, X. C. Pan, H. M. Liu, X. H. Niu, Y. X. Wang, D. F. Xu, H. F. Yang, B. P. Xie, F. Q. Song, P. Dudin, T. K. Kim, M. Hoesch, P. Kumar Das, I. Vobornik, X. G. Wan, and D. L. Feng, Signature of strong spin-orbital coupling in the large nonsaturating magnetoresistance material WTe$_2$, Phys. Rev. Lett. **115**, 166601 (2015).

[12] M. Singh, P. Saha, V. Nagpal, and S. Patnaik, Superconductivity and weak anti-localization in nodal-line semimetal SnTaS$_2$, Supercond. Sci. Technol. **35**, 084003 (2022).

[13] Z. Hao, W. Chen, Y. Wang, J. Li, X.-M. Ma, Y.-J. Hao, R. Lu, Z. Shen, Z. Jiang, W. Liu, Q. Jiang, Y. Yang, X. Lei, L. Wang, Y. Fu, L. Zhou, L. Huang, Z. Liu, M. Ye, D. Shen, J. Mei, H. He, C. Liu, K. Deng, C. Liu, Q. Liu, and C. Chen, Multiple Dirac nodal lines in an in-plane anisotropic semimetal TaNiTe$_5$, Phys. Rev. B **104**, 115158 (2021).

[14] R. Ye, T. Gao, H. Li, X. Liang, and G. Cao, Anisotropic giant magnetoresistance and de Hass-van Alphen oscillations in layered topological semimetal crystals, AIP Advances **12**, 045104 (2022).

[15] D. Wang, F. Tang, J. Ji, W. Zhang, A. Vishwanath, H. C. Po, and X. G. Wan, Two-dimensional topological materials discovery by symmetry-indicator method, Phys. Rev. B **100**, 195108 (2019).

[16] M. Ashton, J. Paul, S. B. Sinnott, and R. G. Hennig, Topology-scaling identification of layered solids and stable exfoliated 2D materials, Phys. Rev. Lett. **118**, 106101 (2017).

[17] C. Xu, Y. Liu, P. Cai, B. Li, W. Jiao, Y. Li, J. Zhang, W. Zhou, B. Qian, X. Jiang, Z. Shi, R. Sankar, J. Zhang, F. Yang, Z. Zhu, P. Biswas, D. Qian, X. Ke, and X. Xu, Anisotropic transport and quantum oscillations in the quasi-one-dimensional TaNiTe$_5$: evidence for the nontrivial band topology, J. Phys. Chem. Lett. **11**, 7782 (2020).

[18] Z. Chen, M. Wu, Y. Zhang, J. Zhang, Y. Nie, Y. Qin, Y. Han, C. Xi, S. Ma, X. Kan, J. Zhou, X. Yang, X. Zhu, W. Ning, and M. Tian, Three-dimensional topological semimetal phase in layered TaNiTe$_5$ probed by quantum oscillations, Phys. Rev. B **103**, 035105 (2021).

[19] E. W. Liimatta and J. A. Ibers, Synthesis, structures, and conductivities of the new layered compounds Ta$_3$Pd$_3$Te$_{14}$ and TaNiTe$_5$. J. Solid State Chem. **78**, 7 （1989）.

[20] S. Hikami, A. Larkin, and Y. Nagaoka, Spin-orbit interaction and magnetoresistance in the two dimensional random system. Prog. Theor. Phys. **63,** 707 (1980).

[21] P. W. Anderson, E. Abrahams, and T. V. Ramakrishnan, Possible explanation of nonlinear conductivity in thin-film metal wires, Phys. Rev. Lett. **43**, 718 (1979).

[22] E. Abrahams, P. W. Anderson, P. A. Lee, and T. V. Ramakrishnan, Phys. Rev. B **24**, 6783 (1981).

[23] J. J. Lin and J. P. Bird, Recent experimental studies of electron dephasing in metal and semiconductor mesoscopic structures, J. Phys.: Condens. Matter **14**, R501 (2002).

[24] O. Pavlosiuk, D. Kaczorowski and P. Wisniewski, Shubnikov-de Haas oscillations, weak antilocalization effect and large linear magnetoresistance in the putative topological superconductor LuPdBi, Sci. Rev. **5**, 9158 (2015).

[25] A. Laha, P. Rambabu, V. Kanchana, L. Petit, Z. Szotek, and Z. Hossain, Experimental and theoretical study of the correlated compound YbCdSn: Evidence for large magnetoresistance and mass enhancement, Phys. Rev. B **102**, 235135 (2020).

[26] S. Mangelsen, P. G. Naumov, O. I. Barkalov, S. A. Medvedev, W. Schnelle, M. Bobnar, S. Mankovsky, S. Polesya, C. Nather, H. Ebert, and W. Bensch, Large nonsaturating magnetoresistance and pressure-induced phase transition in the layered semimetal HfTe$_2$, Phys. Rev. B **96**, 205148 (2017).

[27] X. Yang, H. Bai, Z. Wang, Y. Li, Q. Chen, J. Chen, Y. Li, C. Feng, Y. Zheng, and Z. Xu, Giant linear magneto-resistance in nonmagnetic PtBi$_2$, Appl. Phys. Lett. **108**, 252401 (2016).

[28] Y. Yan, L.-X. Wang, D.-P. Yu, and Z.-M. Liao, Large magnetoresistance in high mobility topological insulator Bi$_2$Se$_3$, Appl. Phys. Lett. **103**, 033106 (2013).

[29] A. B. Pippard, *Magnetoresistance in Metals*, (Cambridge University Press. Cambridge, 2009).

[30] L. Wang, I. Gutierrez-Lezama, C. Barreteau, N. Ubrig, E. Giannini, and A. F. Morpurgo, Tuning magnetotransport in a compensated semimetal at the atomic scale, Nat. Commun. **6**, 8892 (2015).

[31] D. V. Khveshchenko, Magnetic-field-induced insulating behavior in highly oriented pyrolitic graphite, Phys. Rev. Lett. **87**, 206401 (2001).

[32] R. Singha, A. Pariari, B. Satpati, and P. Mandal, Magnetotransport properties and evidence of a topological insulating state in LaSbTe, Phys. Rev. B **96**, 245138 (2017).

[33] Y. Zhou, Z. Lou, S. Zhang, H. Chen, Q. Chen, B. Xu, J. Du, J. Yang, H. Wang, C. Xi, L. Pi, Q. Wu, O. V. Yazyev, and M. Fang, Linear and quadratic magnetoresistance in the semimetal SiP$_2$, Phys. Rev. B **102**, 115145 (2020).

[34] Y. L. Wang, L. R. Thoutam, Z. L. Xiao, J. Hu, S. Das, Z. Q. Mao, J. Wei, R. Divan, A. Luican-Mayer, G. W. Crabtree, and W. K. Kwok, Origin of the turn-on temperature behavior in WTe$_2$, Phys. Rev. B **92**, 180402(R) (2015).

[35] Q. L. Pei, W. J. Meng, X. Luo, H. Y. Lv, F. C. Chen, W. J. Lu, Y. Y. Han, P. Tong, W. H. Song, Y. B. Hou, Q. Y. Lu, and Y. P. Sun, Origin of the turn-on phenomenon in $T_d$-MoTe$_2$, Phys. Rev. B **96**, 075132 (2017).

[36] M. K. Chan, M. J. Veit, C. J. Dorow, Y. Ge, Y. Li, W. Tabis, Y. Tang, X. Zhao, N. Barisic, and M. Greven, In-plane magnetoresistance obeys Kohler's rule in the pseudogap phase of cuprate superconductors, Phys. Rev. Lett. **113**, 177005 (2014).

[37] G. Kastrinakis, Paramagnons, weak disorder and positive giant magnetoresistance, Europhys. Lett. **42**, 345 (1998).

[38] G. Kastrinakis, Theoretical model for the extreme positive magnetoresistance, arXiv:2306.07020.

[39] W. Zhu, C. Song, L. Han, T. Guo, H. Bai, and F. Pan, Van der Waals lattice-induced colossal magnetoresistance in $Cr_2Ge_2Te_6$ thin flakes, Nat. Commun. **13**, 6428 (2022).

[40] D.-Y. Wang, Q. Jiang, K. Kuroda, K. Kawaguchi, A. Harasawa, K. Yaji , A. Ernst, H.-J. Qian, W.-J. Liu, H.-M. Zha, Z.-C. Jiang, N. Ma, H.-P. Mei, A. Li, Coexistence of strong and weak topological orders in a quasi-one-dimensional material, Phys. Rev. Lett. **129**, 146401 (2022).

[41] L. R. Thoutam, Y. L. Wang, Z. L. Xiao, S. Das, A. Luican-Mayer, R. Divan, G.W. Crabtree, and W. K. Kwok, Temperature-dependent three-dimensional anisotropy of the magnetoresistance in $WTe_2$, Phys. Rev. Lett. **115**, 046602 (2015).

[42] G. Blatter, V B. Geshkenbein, and A. I. Larkin, From isotropic to anisotropic superconductors: A scaling approach, Phys. Rev. Lett. **68**, 875 (1992).

[43] I. Pletikosic, M. N. Ali, A. V. Fedorov, R. J. Cava, and T. Valla, Electronic structure basis for the extraordinary magnetoresistance in $WTe_2$, Phys. Rev. Lett. **113**, 216601 (2014).

[44] K. Noto and T. Tsuzuku, A simple two-band theory of galvanomagnetic effects in graphite in relation to the magnetic field azimuth, Jpn. J. Appl. Phys. **14**, 46 (1975).

[45] T. Ishida, K. Inoue, K. Okuda, H. Asaoka, Y. Kazumata, K. Noda, and H. Takei, Anisotropy of superconductivity in an untwined $YBa_2Cu_3O_7$ single crystal, Phys. C: Superconductivity **263**, 260 (1996).

[46] N. Kumar, C. Shekhar, M. Wang, Y. Chen, H. Borrmann, and C. Felser, Large out-of-plane and linear in-plane magnetoresistance in layered hafnium pentatelluride, Phys. Rev. B **95**, 155128 (2017).

[47] Z. Hou, B. Yang, Y. Wang, B. Ding, X. Zhang, Y. Yao, E. Liu, X. Xi, G. Wu, Z. Zeng, Z. Liu, and W. Wang, Large and anisotropic linear magnetoresistance in single crystals of black phosphorus arising from mobility fluctuations, Sci. Rep. **6**, 23807 (2016).

[48] X. Zhang, V. Kakani , J. M. Woods, J. J. Cha , and X. Shi, Thickness dependence of magnetotransport properties of tungsten ditelluride, Phys. Rev. B **104**, 165126 (2021).

[49] A. E. M. Smink, J. C. de Boer, M. P. Stehno, A. Brinkman, W. G. van der Wiel, and H. Hilgenkamp, Gate-tunable band structure of the $LaAlO_3$-$SrTiO_3$ interface, Phys. Rev. Lett. **118**, 106401 (2017).

[50] S. Gu, K. Fan, Y. Yang, H. Wang, Y. Li, F. Qu, G. Liu, Z.-A. Li, Z. Wang, Y. Yao, J. Li, L. Lu, and F. Yang, Classical linear magnetoresistance in exfoliated $NbTe_2$ nanoflakes, Phys. Rev. B **104**, 115203 (2021).

[51] S. C. Bodepudi, X. Wang, A. P. Singh, and S. Pramanik, Thickness dependent interlayer magnetoresistance in multilayer graphene stacks, J. Nanomater. **2016**, 8163742 (2016).

[52] R. S. Singh, X. Wang, W. Chen, A. Ariando, and A. T. S. Wee, Large room-temperature quantum linear magnetoresistance in multilayered epitaxial graphene: Evidence for two-dimensional magnetotransport, Appl. Phys. Lett. **101**, 183105 (2012).

**Figure captions**

**Figure 1** (a) X-ray diffraction pattern of TaNiTe$_5$ single crystal. (b) High-resolution transmission electron microscopy image. (c) Energy dispersive X-ray spectroscopy images of three elements Ta, Ni, and Te.

**Figure 2** Resistivity $\rho_{xx}$ as a function of temperature $T$ at various magnetic fields $B$s for samples (a) S1, (b) S3, (c) S4, and (d) S5. Dashed lines are fits according to Fermi liquid model $\rho_{xx} = A + CT^2$. Insets are optical microscope images of devices. The current directions are marked by arrows.

**Figure 3** (a) Magnetoresistance (MR) at 2 K for five samples. (b) MR at various $T$s for sample S3. Inset shows the MR as a function of $T$ for a given field $B = 8\,\mathrm{T}$. (c) Magnetoconductivity curves at various $T$s for sample S3. Solid lines are fits according to Eq. (1). (d) The extracted parameters $\alpha$ and $L_\varphi$ as a function of $T$. Solid line is a fit.

**Figure 4** (a) Hall resistivity $\rho_{xy}$ as a function of $B$ at various T for sample S4. Solid lines are fits with Eq. (3). (b) The fit-obtained carrier mobilities as a function of $T$. (c) The fit-obtained carrier concentrations as a function of $T$. Inset is a schematic diagram of the band structure near the Fermi level. (d) MR at 2 K for sample S3. Solid line is calculated result according to Eq. (2).

**Figure 5** Kohler's rule by plotting MR versus $B/\rho_0$ from 2 to 200 K for samples (a) S3, (b) S4, and (c) S5. Red dashed lines are fits according to Eq. (4). Inset in (a) shows the normalized MR with value at 2 K for sample S3. (d) $\rho_{xx}$ as a function of $T$ at various $B$s for sample S3. Solid lines are fits with Eq. (5).

**Figure 6** (a) The $ac$ and (b) $ab$ planes of TaNiTe$_5$ crystal structure. The yellow rectangles in (a) indicate the one-dimensional NiTe$_2$ chains.

**Figure 7** (a) MR at 2 K for sample S3 when $B$ is tilted along various directions. Inset is an enlargement of MR curve when $B$ is parallel to the $c$ axis. (b) The same data in (a) with $B$ multiplied by a factor $\varepsilon_\theta$. Inset is a sketch of experimental configuration. (c) $\varepsilon_\theta$ vs tilted angle $\theta$ at various $T$s. Solid lines are fits with Eq. (7). (d) $T$ dependence of the extracted mass anisotropy $\gamma$ for sample S3. Inset shows $\gamma$ as a function of the thickness of nanoflake at 2 K.

**Table 1:** Electrical parameters of the exfoliated nanoflakes. *RRR* is residual resistivity ratio. MR is magnetoresistance at 2 K and 8 T.

| Sample | thickness (nm) | *RRR* | Current direction | MR (%) |
|--------|----------------|-------|-------------------|--------|
| S1 | 157.6 | 29.1 | *a* axis | 29.2 |
| S2 | 48.3 | 25.7 | *a* axis | 43.0 |
| S3 | 88.6 | 25.8 | *c* axis | 486.8 |
| S4 | 65.7 | 16.3 | *c* axis | 408.3 |
| S5 | 33.7 | 28.3 | *c* axis | 132.5 |

Figure 1

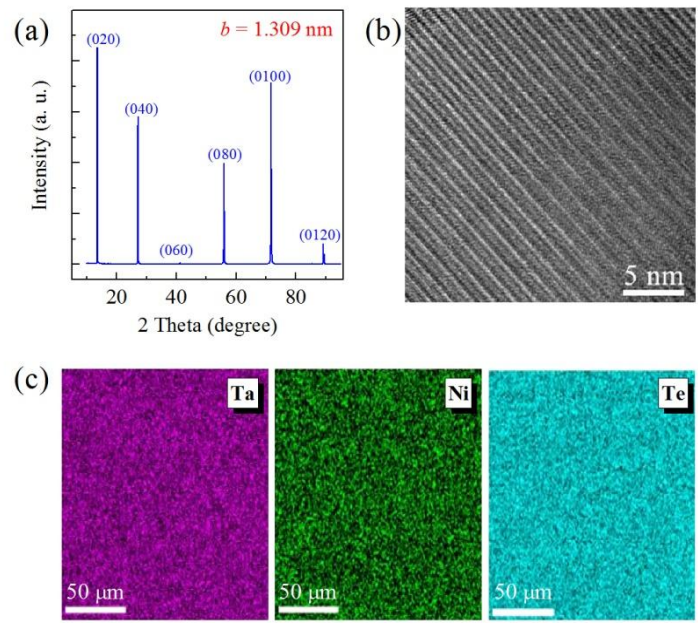

Figure 2

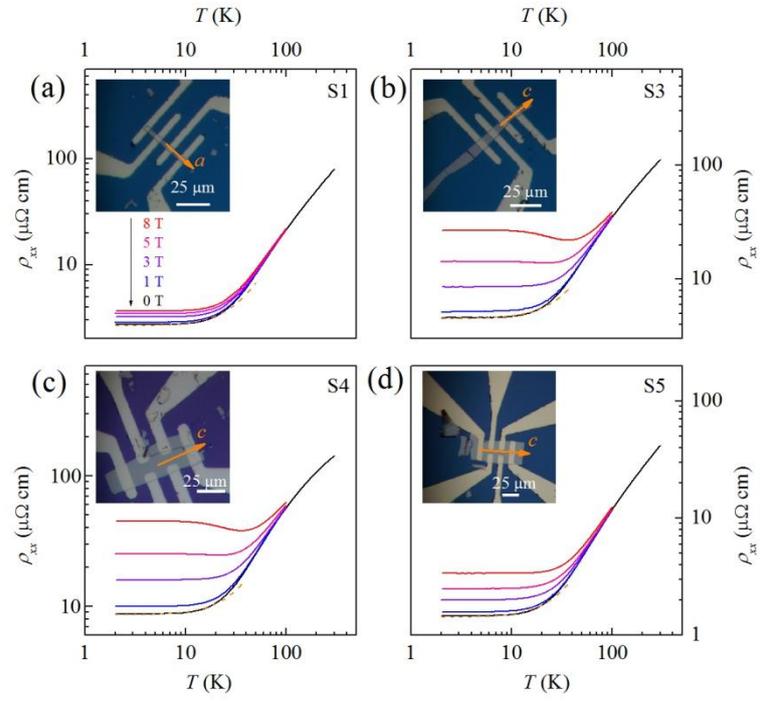

Figure 3

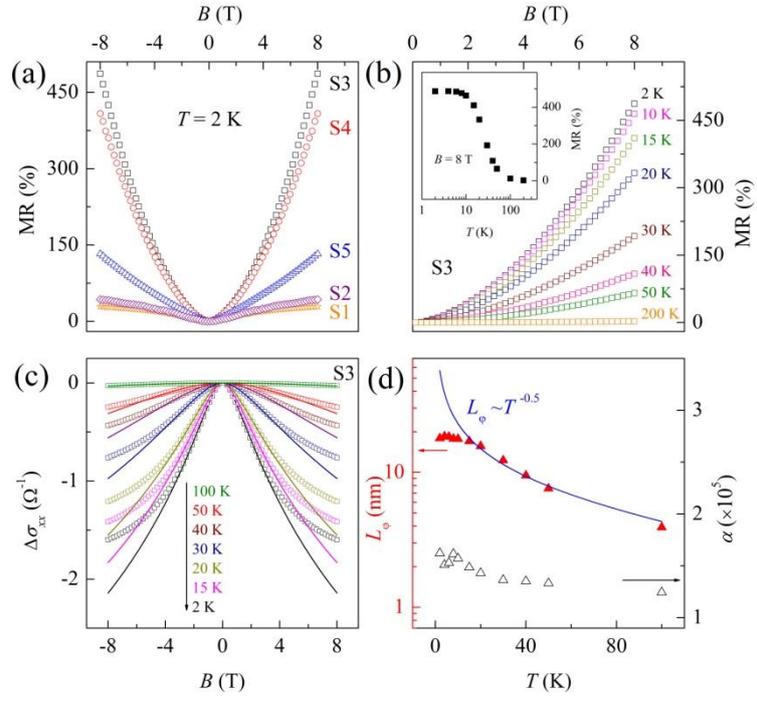

Figure 4

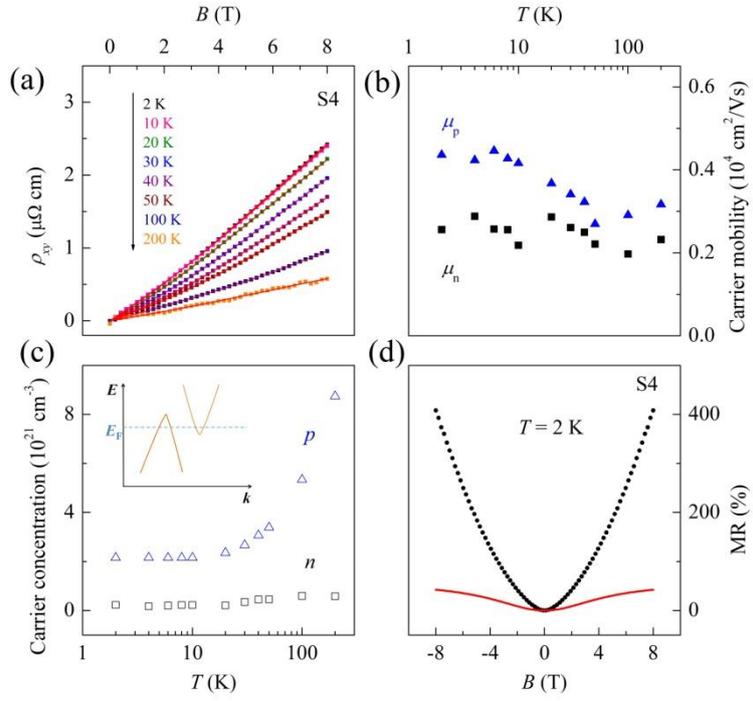

Figure 5

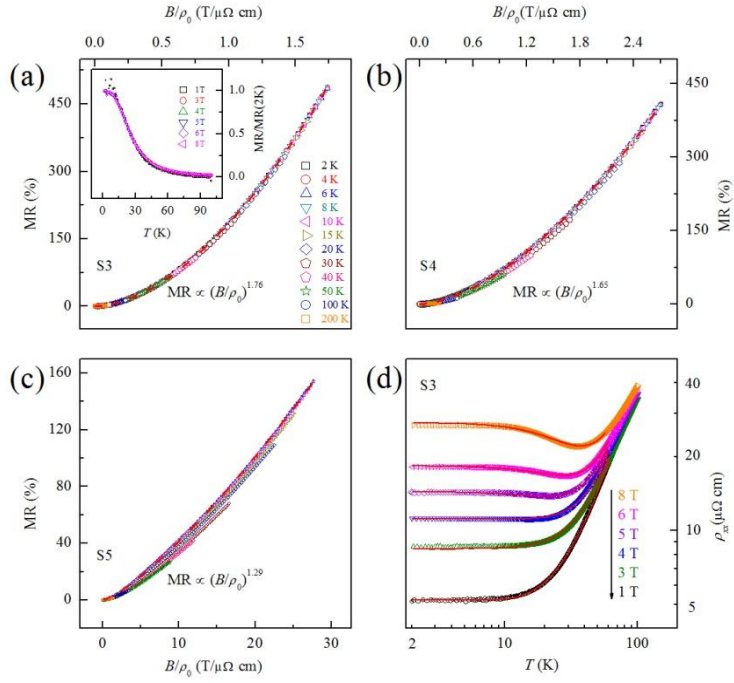

Figure 6

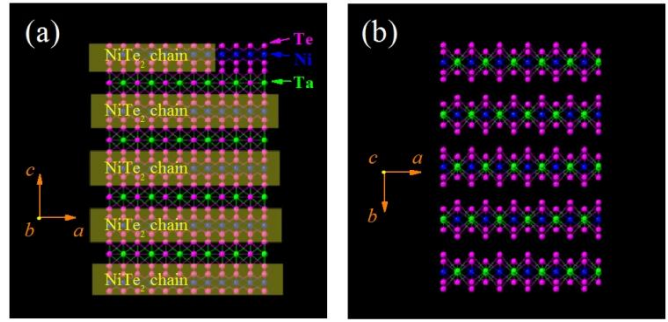

Figure 7

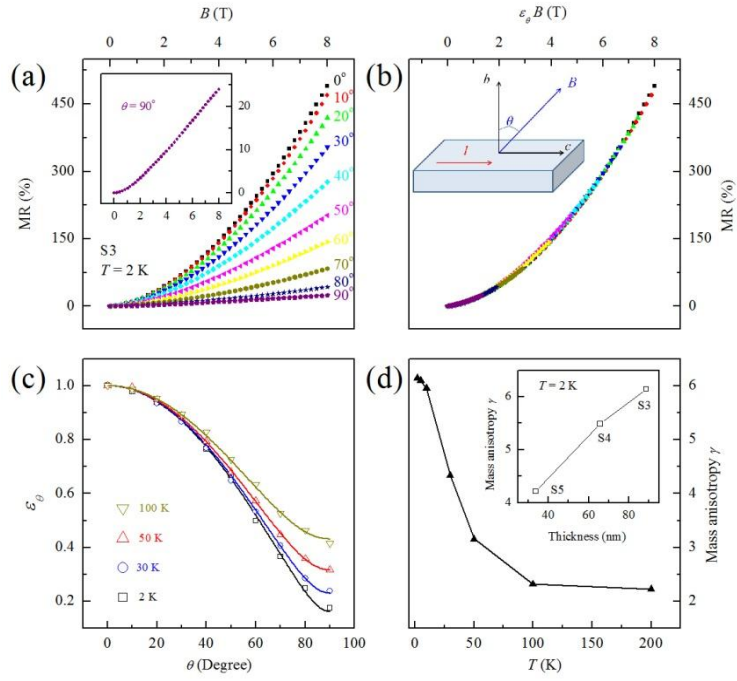